\documentclass[11pt,twoside]{article}

\usepackage{asp2006}

\usepackage{epsf}

\usepackage{graphicx}
\usepackage{lscape}

\begin{document}

\title{The Horsehead nebula, a template source for interstellar physics and chemistry}   

\author{Maryvonne Gerin\altaffilmark{1}, J\'er\^ome Pety\altaffilmark{2,1} and
Javier R. Goicoechea\altaffilmark{3} }   

\affil{$^1$LERMA-LRA, CNRS UMR8112, Observatoire de Paris and Ecole Normale
  Sup\'erieure, 24 Rue Lhomond, 75231 Paris cedex 05, France.
email : maryvonne.gerin@ens.fr}    

\affil{$^2$IRAM, 300 rue de la piscine, 38046 Grenoble, France. email:pety@iram.fr}

\affil{$^3$Laboratorio de Astrof\'{\i}sica Molecular. Centro de Astrobiolog\'{\i}a. CSIC-INTA.
Carretera de Ajalvir, Km 4. Torrej\'on de Ardoz, 28850, Madrid, Spain. 
e-mail: goicoechea@damir.iem.csic.es}

\begin{abstract} 
We present a summary of our previous investigations of the
physical and chemical structure of the Horsehead nebula, and discuss
how these studies led to advances on the understanding of the impact 
of FUV radiation on the structure of dense interstellar clouds. 
Specific molecular tracers can be used to isolate different environments,
that are more sensitive to changes in the FUV radiation or density than the
classical tracers of  molecular gas : the CO isotopologues or the dust
(sub)millimeter continuum emission. They include the HCO or CCH radicals for
the FUV illuminated interfaces, or the molecular ions H$^{13}$CO$^+$, DCO$^+$
and other deuterated species (DNC, DCN) for the cold dense core.
We discuss future prospects in the context of Herschel and ALMA.
\end{abstract}









\section{Introduction} 
In this symposium, we have learned how the progresses in detector
technologies, with the development of Schottky mixers, InSb hot electron
bolometers and finally SIS mixers led to spectacular discoveries on the
chemical diversity and physical properties of the interstellar medium.
The role of the  interstellar medium in the life cycle of baryonic matter
has been gradually uncovered in the last century. Following the discovery
of the metallic lines in the spectra of reddened stars, the first molecules
were recognized in the late 1930's. As these spectral features were extremely
narrow, with velocity dispersions of a few kms$^{-1}$ at most, 
the perspectives of
radioastronomy and heterodyne spectroscopy for studying the 
interstellar medium, and
especially its coldest and densest regions was quickly recognized. 
Molecular lines offer 
a wealth of diagnostics of the physical conditions, the chemical composition
and the gas dynamics from the regions where they are formed.
Molecular clouds themselves play a special role in the life cycle of
interstellar matter, as the birth places of the new generation of stars 
but also as the main sites for interaction of stars with their environment.
Stars affect their environment through their radiative (through their X and UV
radiation), mechanical (winds, shocks) and chemical (nucleosynthesis)  
feedback. The understanding of the properties of molecular clouds, 
their structure and their
relation to the formation of stars has triggered the development of
millimeter and submillimeter astronomy. Measuring the cooling radiation
from molecular clouds  is a key issue that determines our ability
to detect the interstellar matter and understand how it can collapse and form
stars. T. Phillips and his co-workers have brought a decisive contribution
to this field, by making pioneering detections of the submillimeter lines
of CO, atomic carbon and hydrides. His prediction of the spectrum
of a typical interstellar cloud \citet{phi87} has been vastly cited 
and reproduced
to illustrate the scientific discovery potential of submillimeter astronomy.

The neutral interstellar gas cools through line emission, the most intense
lines being the fine structure lines of ionized carbon, atomic oxygen and
carbon, together with the rotational lines of CO and water. Molecular
hydrogen (H$_2$) itself can be an important coolant, especially in dense
photon-dissociation regions (PDRs) and shocks.  
Averaged over the whole interstellar medium of a galaxy, this line  cooling
radiation from PDRs makes a significant contribution to the total line
emission at large scales \citep{phi81,phi87,wol08,cubik}. 
\citet{keene} have pioneered this field by
mapping the atomic carbon line across the edge of two well known PDRs, S~140
and M~17.  They showed that the atomic carbon line emission is very extended,
and does not show the expected spatial pattern for a uniform density, plane 
parallel cloud.
They discussed how the knowledge of the source geometry and spatial structure
 are particularly important for understanding their line emission. 

\section{Physical structure of the Horsehead nebula}
\subsection{Summary of physical properties}

The Horsehead nebula is one of the most well known example of a molecular
cloud. It appears as a dark patch against the bright emission from the HII
region IC~434 at visible wavelengths, and as a bright line and
continuum emission source at IR and (sub)millimeter wavelengths. 
Following the mapping of the CO(3--2) line emission with the 
CSO\footnote{see the CSO www page at www.submm.caltech.edu/}, we
have started investigating the physical and chemical structure of the
Horsehead nebula. The source was chosen as it is a good template of 
photon dissociation region (PDR), where molecular gas is heated and
photodissociated by Far UV (FUV; $6$eV $< h \nu < 13.6$eV) radiation from 
nearby massive stars. 
The Horsehead nebula is  particularly interesting  as its
geometry is simple. The illuminated edge is viewed very close to edge-on
and the illuminating star ($\sigma$ Ori) lies at the same distance. 
The properties of the Horsehead nebula are summarized in Table \ref{tab:id}.

\begin{figure}[!ht]
\plotfiddle{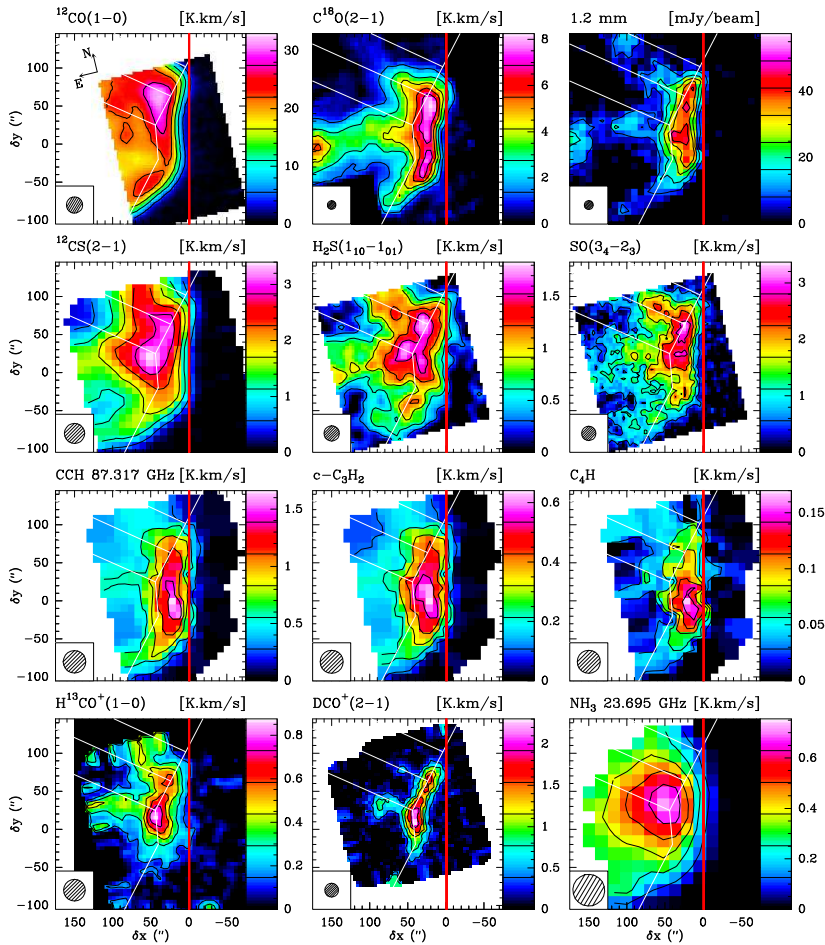}{11.5cm}{270}{80}{80}{-330}{410}
\label{fig:30m}
\caption{Display of the maps obtained with the IRAM 30m and Effelsberg
 (for NH$_3$) 
telescopes towards the Horsehead nebula. All line maps show the integrated
intensity in K.kms$^{-1}$ between 9.0 and 12.0 kms$^{-1}$. The levels and
color scale are shown on the right side of each map. The telescope beam is
shown as a grey circle in the lower left corner. All maps have been rotated by
14$^\circ$ counter-clock wise in order to display the PDR edge vertically. This
edge is shown as the red vertical line. The white lines are drawn to help 
localize the main structures. }
\end{figure}

\begin{table}[ht!]
\caption{\label{tab:id}Summary of the physical properties of the Horsehead
  nebula}
\smallskip
\begin{center}
{\small
\begin{tabular}{ll}
\tableline
\noalign{\smallskip}
Name & Barnard 33 (Horsehead nebula)\\
RA, DEC (2000) & 05h40m54.27s, $-02^\circ 28' 00"$ \\
Distance from Earth & $\sim 400$ pc (1'' = 0.002 pc) \\
Illuminating Star & $\sigma$Ori (O9.5V) at 0.5$^\circ$ (3.5 pc), PA 76$^\circ$ \\
FUV Radiation field & 60 (Draine unit) \\
Depth along the line of sight & $\sim 0.1$ pc \\
Inclination of the line of sight & $\le 6^\circ$ \\
Density profile &  Steep gradient \\ 
                & $n \sim r^{-3}$ from $\sim 10^5$ cm$^{-3}$
 to 10$^3$ cm$^{-3}$ in 10''\\
& and lower density halo $5 - 10 \times 10^3 $ cm$^{-3}$\\
Kinetic temperature & from T$_K$ $\sim 15$ K in the cold core \\
                    & up to T$_K$ $\sim 100 $K at the edge\\

Thermal pressure & $\sim 4 \times 10^6$ Kcm$^{-3}$ \\
\noalign{\smallskip}
\tableline
\end{tabular}}
\end{center}
\end{table}

Figures 1 and 2 present the line integrated
intensities of the interstellar molecules we have observed so far (\cite{tey04,pet05,pet07,goi06,goi09,ger09}), together
with the excited vibrational line of H$_2$ at 2.1$\mu$m from \cite{hab05}, and
the ISOCAM 7.7$\mu$m map from \cite{abe03}.

\begin{figure}[!ht]
\plotfiddle{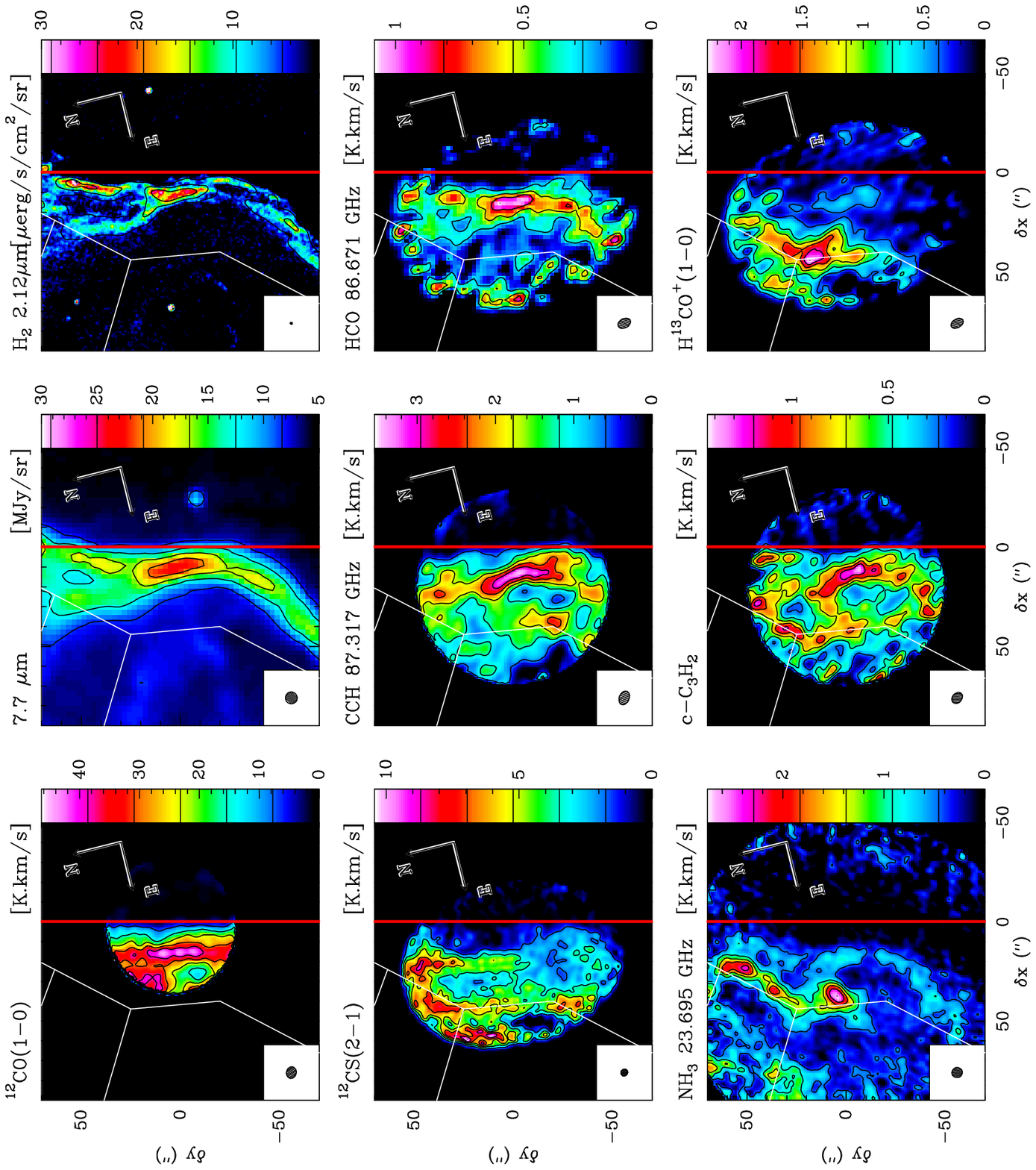}{12cm}{270}{75}{75}{-310}{400}
\label{fig:pdbi}
\caption{Display of the images obtained with the IRAM PdBI and NRAO VLA (for
  NH$_3$) towards the
Horsehead nebula. The H$_2$ 2.1 $\mu$m image is from \cite{hab05} and
the ISOCAM 7.7 $\mu$m image from \cite{abe03}. Short spacings from the
IRAM-30m (or Effelsberg 100m for NH$_3$) have been combined with the interferometer
data to restore images with all spatial scales. The levels and
color scale are shown on the right side of each map. The telescope beam is
shown as a grey circle in the lower left corner. All maps have been rotated by
14$^\circ$ counter-clock wise in order to display the PDR edge vertically. This
edge is shown as the red vertical line. The white lines are drawn to help 
localize the main structures. }
\end{figure}

\subsection{Electron abundance}
The electron abundance ([e$^-$] = $n_e/n_H$) plays a fundamental
role in the chemistry and dynamics of interstellar gas. It
 can not be directly
measured in FUV shielded gas but can be derived from abundances 
and abundance ratios of key
molecular species, which are sensitive to the presence of electrons. 
Stable molecular ions such as DCO$^+$
and HCO$^+$ have been traditionally used to estimate the
electron abundance in cold cores  because (i) they
are abundant and easily observable (ii) dissociative recombination
is their main destruction route, and thus their
abundances are roughly inversely proportional to the electron
abundance (e.g. \citet{gue82,wot82,mar07,hez08}).
 On the other
hand, the presence of reactive ions (species such as HOC$^+$ or
CO$^+$ that react rapidly with H$_2$) indicates
 high electron abundance regions, e.g., the UV irradiated
cloud surfaces (e.g., \citet{smi02,fue03}).
The electron abundance is predicted to steeply decrease in a PDR from 
$\sim 10^{-4}$ at the cloud edge where the FUV radiation field is intense,
down to $< 10^{-7}$ in the cold and well shielded gas. The magnitude and
shape of this gradient is sensitive to the elemental abundances, especially
sulfur, to the cosmic ray ionization rate, and to the efficiency of
recombination reactions and electron attachment processes with PAHs and grains 
\citep{flo07,wak08,goi09}.

By combining pointed observations and maps of H$^{13}$CO$^+$, DCO$^+$,
HOC$^+$ and CO$^+$, \citet{goi09} have investigated the electron abundance
gradient across the edge of the Horsehead nebula. They showed that a 
consistent fit of both the PDR edge and the adjacent cold core can
be obtained with a cosmic ray ionization rate of
$\zeta = 5 \pm 3 \times 10^{-17}$ s$^{-1}$, assuming a low
metal abundance of $[M^+] = 10 ^{-9}$. 
The electron abundance in the Horsehead edge follows a steep gradient,
    with a scale length of $\sim$0.05\,pc (or $\sim$25$''$), from
    [e$^-$]$\simeq$10$^{-4}$   
times $\sim$10$^{-9}$ in the core. Low metal abundances are
required in the core,  namely $[M^+] < 4\times 10^{-9}$ 
(with respect to H nuclei), to enable a significant deuterium
fractionation of HCO$^+$. This implies a very significant depletion of
metals onto dust grains as the solar abundance of metals is 
$[M^+] \simeq 8.5 \times 10^{-5}$; \citep{and89}. 
The inclusion of  PAHs in the chemical model modifies the electron abundance
 gradient and decreases the metal depletion required in pure gas-phase 
models as a good agreement is obtained with metal abundance 
$[M^+] = 3\pm1 \times 10^{-6}$ for a standard PAH abundance of 10$^{-7}$.
    Indeed, PAH$^-$ anions acquire large abundances, being more
    efficient in neutralizing atomic ions than electrons.
PAHs have a significant impact on the negative charge budget  of cold gas if
their abundance is significant ([PAH]$>$10$^{-8}$).

\section{Chemical gradients}
\subsection{Determination of chemical abundances}
Molecular line emission does not directly probe the abundances
of a given species, even when several lines of the same species are detected.
 A minimum treatment of the molecular excitation and line radiative transfer
is often needed to distinguish ``excitation gradients'' from true ``chemical
variations''. Our strategy is always to observe several rotational lines of
the same species (sometimes with line maps, sometimes using high signal to
noise ratio spectra at representative positions). In addition, we also try to
observe molecular isotopologues, if possible, when the main line opacities can
complicate the interpretation and lead to large errors on the derived
abundance. Molecular excitation in PDRs is always far from LTE and
different collisional and radiative effects compete at rates that depend on
the physical conditions and on the molecule under study. In the last years, we
have always tried to consistently link the observed line intensities with
the abundances predicted by our chemical models by detailed non-LTE
excitation and radiative transfer modeling adapted to the Horsehead geometry
(see Appendix in \citet{goi06}).
 
Table \ref{tab:chem1} and \ref{tab:chem2}
list the molecular abundances for the PDR region and the dense core
respectively, derived following the method described above. 

\subsection{Observed chemical abundances}

It is obvious, when comparing the maps of various molecules
shown in Fig. 1 and 2 that the molecular abundances are
not uniform across the Horsehead nebula. Molecular species like
HCO and CCH closely delineate the western illuminated edge of the nebula. 
On the contrary, H$^{13}$CO$^+$ and the deuterated species DCO$^+$ and DNC are
mostly located in the dense cold core. Finally the dust continuum emission,
as well as C$^{18}$O trace the bulk of the core, but do not show a sharp
peak associated with the dense core.
These maps have been discussed in detailed, and  molecular
abundances in our previous papers.

\begin{table}[ht!]
\caption{\label{tab:chem1}Summary of measured molecular column densities 
and abundances in the PDR. Offsets refer to
 RA(2000) = 05:40:54.27, DEC(2000) = -02:28:00}
\smallskip
\begin{center}
{\small
\begin{tabular}{l c c c c l}
\tableline
\noalign{\smallskip}
Species & HPBW & Column density & Abundance & Reference & Offsets\\
        & [arcsec] &      [cm$^{-2}$]   & ${N(X)}\over{N(H)+2N(H_2)}$& &
        [arcsec]\\ 
\noalign{\smallskip}
\tableline
\noalign{\smallskip}
H$_2$ & 12 & $1.9 \times 10^{22}$ & 0.5 & 3,4,5,6& (-5,0)\\ 
C$^{18}$O & $6.5 \times 4.3$  &$4.0 \pm 0.5 \times 10^{15}$ & $1.9 \times 10^{-7}$&  2 & (-6,-4)\\
CCH & $7.2 \times 5.0$ & $3.0 \pm 0.5 \times 10^{14}$ & $1.4 \times 10^{-8}$& 2 & (-6,-4) \\
c-C$_3$H$_2$ & $6.1 \times 4.7$ & $2.4 \pm 1.0 \times  10^{13}$ & $1.1 \times
10^{-9}$& 2 & (-6,-4) \\
c-C$_3$H$_2$ & 29 & $9.3 \pm 0.2 \times 10^{12}$ & $6.5 \times 10^{-10}$ & 1 & (-10,0)\\
l-C$_3$H$_2$ & 27 & $< 3.3 \times 10^{11}$ & $ < 4.6 \times 10^{-11}$ & 1,7 & (-10,0)\\
c-C$_3$H & 28 & $3.9 \pm 0.5 \times 10^{12}$ & $2.7 \times 10^{-10}$ & 1 & (-10,0)\\
l-C$_3$H & 28 & $2.1 \pm 0.7 \times 10^{12}$ & $1.4 \times 10^{-10}$ & 1 & (-10,0 \\
C$_4$H & $6.1 \times 4.7$ & $4.0 \pm 1.0 \times 10^{13}$ & $1.0 \times 10^{-9}$& 2 &
(-6,-4) \\
C$_6$H & 28 & $9 \pm 1 \times 10^{11}$ & $2.2 \times 10^{-11}$ & 8 & (-6,4) \\
CS & 10 & $8.1 \pm 1.0 \times 10^{13}$ & $2.0 \times 10^{-9}$ & 3 & (+4,0)\\
C$^{34}$S & 16 & $3.7 \pm 0.5 \times 10^{12}$ & $9.2 \times 10^{-11}$ & 3 & (+4,0)\\
HCS$^+$ & 29 & $6.8 \pm 0.5 \times 10^{11}$ & $1.7 \times 10^{-11}$ & 3 & (+4,0)\\
H$^{13}$CO$^+$ & 14 & $5.8 \pm 1 \times  10^{11}$ & $1.5 \times 10^{-11}$ & 5, 6 & (-5,0)\\
HOC$^+$ & 27.5 & $1.8 \pm 0.6 \times 10^{11}$ & $ 4.0 \times 10^{-12}$ & 6 & (-5,0)\\
CO$^+$ & 10.4 & $< 1.6 \times 10^{10}$ & $< 2.7 \times 10^{-13}$ & 6 & (-5,0)\\
HCO & 14 & $3.2 \pm 0.6 \times 10^{13}$ & $8.4  \times 10^{-10}$ & 5 & (-5,0)\\
\noalign{\smallskip}
\tableline
\end{tabular}
\smallskip
\par
1 : \citet {tey04}; 2 : \citet{pet05}, 3 : \citet{goi06}, 4 : \citet{pet07},
5 : \citet{ger09}, 6 : \citet{goi09}, 7 : \citet{tey05}, 8 : \citet{agu08}
}
\end{center}
\end{table}

\begin{table}
\caption{\label{tab:chem2}Summary of measured molecular column densities 
and abundances towards the cold dense core. Offsets refer to
 RA(2000) = 05:40:54.27, DEC(2000) = -02:28:00  }
\smallskip
\begin{center}
{\small
\begin{tabular}{lccccl}
\tableline
\noalign{\smallskip}
Species & HPBW & Column density & Abundance & Reference & Offsets\\
        & [arcsec] &      [cm$^{-2}$]   & ${N(X)}\over{N(H)+2N(H_2)}$ & & [arcsec]\\ 
\noalign{\smallskip}
\tableline
\noalign{\smallskip}
H$_2$ & 12 & $2.9\times 10^{22}$ & 0.5 & 6 & (+20,+22)\\
CS & 10 & $1.2 \pm 1.0 \times 10^{14}$ & $2.0 \times 10^{-9}$  & 3 & (+21,+15)\\
C$^{34}$S & 16 & $5.3 \pm 0.5\times 10^{12}$ & $9.1 \times 10^{-11}$  & 3 &(+21,+15)\\
HCS$^+$ & 29 & $6.8 \pm 0.5\times 10^{11}$ & $1.2 \times 10^{-11}$  & 3 &(+21,+15)\\
H$^{13}$CO$^+$ & $6.7 \times 4.4$ & $6.5 \pm 2.5  \times 10^{12}$ & $6.5
\times 10^{-11}$ & 4,6 & (+20,+22)\\
DCO$^+$ & 12 & $7.5 \pm 2.5 \times 10^{12}$ & $8.0 \times 10^{-11}$ & 4,6 & (+20,+22)\\
HCO & 14 & $4.6 \pm 1.0 \times 10^{12}$ & $8.0 \times 10^{-11}$ & 5 & 
(+20,+22)\\
\noalign{\smallskip}
\tableline
\end{tabular}
\smallskip
\par
1 : \citet{tey04}; 2 : \citet{pet05}, 3 : \citet{goi06}, 4 : \citet{pet07},
5 : \citet{ger09}, 6 : \citet{goi09}
}
\end{center}
\end{table}

\begin{figure}[!ht]
\plotone{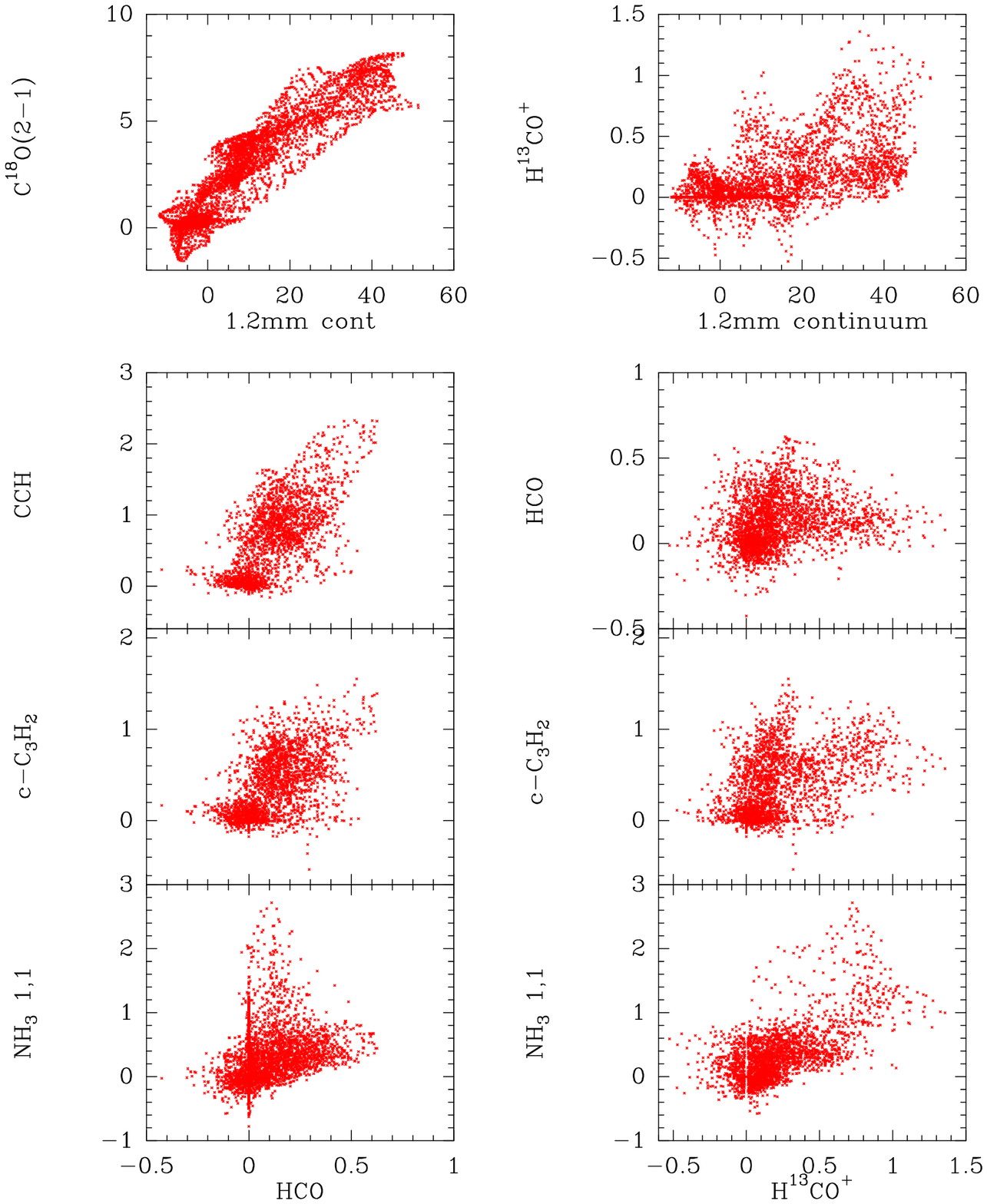}
\label{fig:correl}
\caption{Scatter diagrams of selected couples of images  
of the Horsehead nebula. All images have been resampled to the same spatial
grid. Top line : The 1.2mm dust continuum versus C$^{18}$O(2-1) and
H$^{13}$CO$^+$(1-0).
Left column, from top to bottom, scatter diagrams of CCH, c-C$_3$H$_2$ and
NH$_3$ with HCO. Right column : scatter diagrams of HCO, c-C$_3$H$_2$ and
NH$_3$ with H$^{13}$CO$^+$(1-0).
 }
\end{figure}

In order to demonstrate quantitatively how these various
species are spatially related, we show in Figure 3
a selection of scatter diagrams. The 1.2mm dust continuum  and C$^{18}$O
emissions are well correlated, while the correlation is less strong
for H$^{13}$CO$^+$. The correlation between
HCO and CCH  is good as well, as both species trace the UV
illuminated matter in the PDR. C$_4$H and CCH are equally well 
correlated, as shown by \cite{tey04,pet05}. The recent confirmation of the
detection of C$_6$H in the Horsehead edge by
\cite{agu08} confirms the high abundance 
of unsaturated hydrocarbons and radicals
in this source. 
By contrast, the H$^{13}$CO$^+$ (1-0) emission 
is tracing the dense cold core. It is loosely correlated with the
dust continuum emission. The correlation of H$^{13}$CO$^+$ with
NH$_3$ is quite good.

The situation can be more complicated for other molecules. For instance,
the c-C$_3$H$_2$ maps  shows two emission regions, the PDR and the
cold dense core. The presence of abundant c-C$_3$H$_2$  in the PDR
is confirmed by the good correlation with HCO. The additional emission source
shows up in the same plot as an upper envelope of points
 with strong c-C$_3$H$_2$ and relatively low HCO 
emission. This is confirmed in the c-C$_3$H$_2$ vs H$^{13}$CO$^+$ scatter
plot, that shows two clouds of points with two different slopes.
Although most of the NH$_3$(1,1) emission is produced in the dense and
shielded region, some faint emission is detected in the PDR. This
behavior shows up in the NH$_3$ vs HCO and NH$_3$ vs H$^{13}$CO$^+$
scatter plots as clouds of points deviating from the general trends.
The ammonia data will be further discussed and analyzed in a forthcoming
paper (Pety et al. 2009, in prep.).

\subsection{Chemical models of the Horsehead nebula}

\begin{figure}[!ht]
\plotfiddle{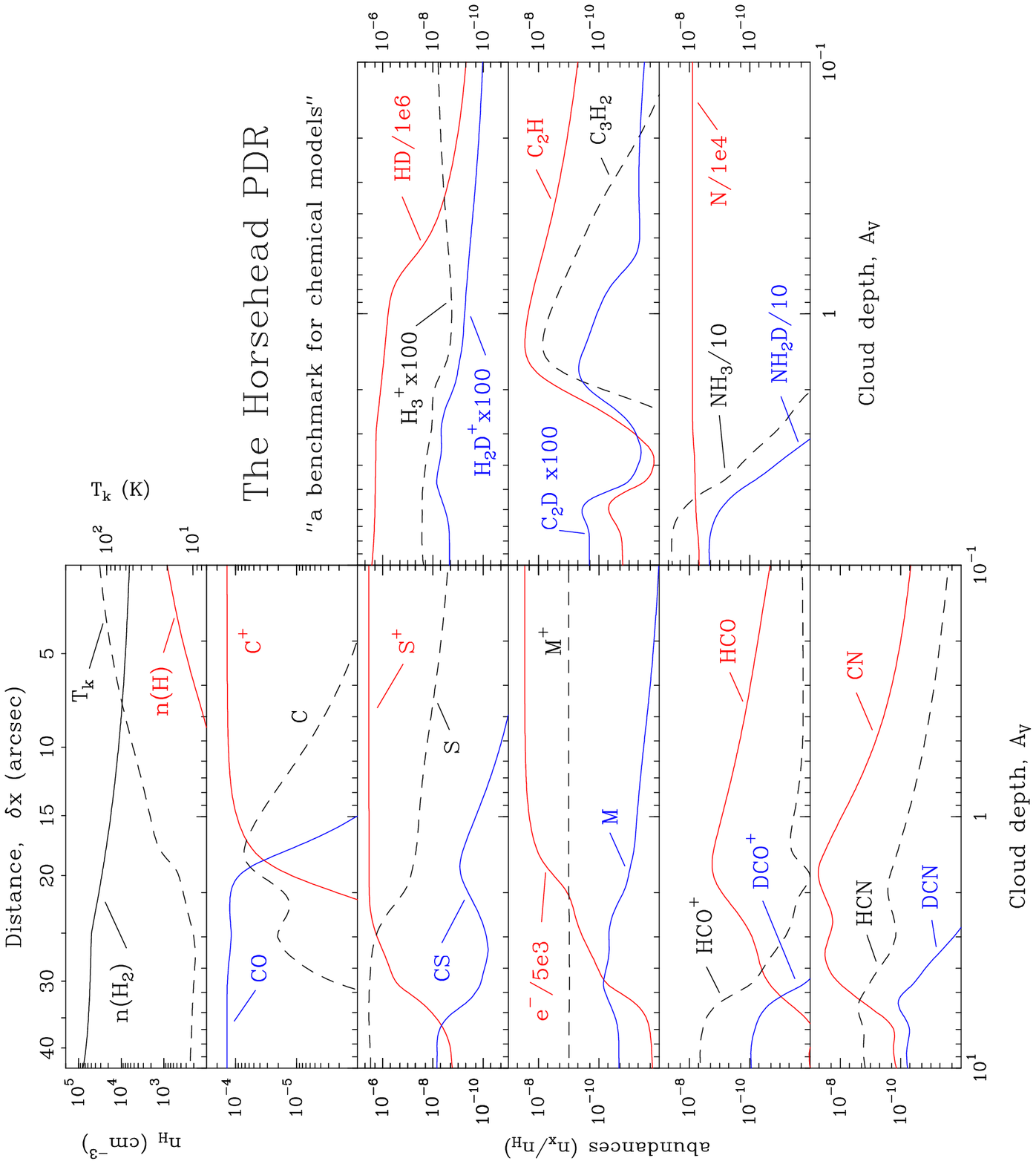}{10.5cm}{270}{65}{65}{-250}{350}
\label{fig:model}
 \caption{
PDR model adapted to the physical conditions in the Horsehead
($\chi$=60 and a density gradient see Goicoechea et al. 2009 for details).
The uppermost panel shows the resulting H$_2$ and H density profiles,
together with the expected gas temperature gradient.
The next panels show the predicted abundances (with
respect to to H nuclei) of the species  we have observed and
analyzed in detail in the past years:
The CO/C/C$^+$ transition (Habart et al. 2005) ,
The low sulfur depletion from CS observations (Goicoechea et al. 2006);
The electron abundance, cosmic-rays ionization-rate and
metal depletion (Goicoechea et al. 2009);
The high  HCO abundance and the HCO/H$^{13}$CO$^+$  ratio
as tracer un FUV-induced chemistry (Gerin et al. 2009);
the chemical deuteration of different molecules towards the cold
and FUV-shielded regions (Pety et al. 2007); the small hydrocarbons
emission in the PDR (Pety et al. 2005); the CN/HCN abundance ratio
and the ammonia emission in the region (Pety et al. 2009 in prep.).}
\end{figure}

Photo-dissociation models have been used to compute the physical
and chemical structure of geometrically simple interstellar clouds, exposed to
FUV radiations. Because of its well known geometry and FUV radiation field,
the Horsehead nebula is a good template source for comparing model predictions
with actual data. These comparisons have been extremely interesting in
identifying the strong points of the present generation of models, and their
remaining weaknesses, and in triggering new improvements.
Figure 4 shows PDR model predictions adapted to
the Horsehead nebula, made with the Meudon PDR model 
\citep{lepetit,goi07,gon08}. Predictions are shown for the CO/C/C$^+$
transition \citep{hab05}, the sulfur chemistry \citep{goi06}, the electron
abundance \citep{goi09}, the deuterium fractionation \citep{pet07}, the
HCO and CCH radicals \citep{pet05,ger09} as well as for the main tracers of
the nitrogen chemistry, CN, HCN, HNC and NH$_3$.  
While the physical structure, and the transition from a FUV dominated
chemistry in the PDR, to a cosmic ray dominated chemistry in the dense core, is
well reproduced by the models, significant discrepancies remain.
The HCO, CCH and C$_4$H radicals are much more abundant in the PDR than
steady state model predictions. \citet{pet05} and \citet{ger09} have suggested
directions for improvements. They include i) the need to include more
neutral-neutral reactions in the chemical networks, that may have a
significant impact on the chemistry (e.g. the O + CH$_2$ reaction in the HCO
case), and ii) the need for considering the coupling of gas phase and solid
phase chemistry as the photodesorption of molecular ices and/or the 
erosion of carbon grains may release significant amounts of radicals in the
gas phase, that can modify current model predictions.

\section{Future prospects with Herschel, NOEMA, and ALMA}

Because of the steep gradients in the physical parameters and chemical
abundances in PDRs such as the Horsehead nebula, an accurate comparison
of observations with models cannot be performed without information on
the spatial variation of the molecular line intensities. Due to the
small spatial scales involved, a combination of
(sub)millimeter interferometer maps with short spacings from single dishes
appears to be the best method for obtaining this spatial information at the
required physical resolution. However, many interesting species show
relatively weak lines, which are bearly detectable with the IRAM PdBI in a
strong and nearby source like the Horsehead nebula (for
instance the C$_4$H lines). To access a more complete sample of molecular
tracers, a more sensitive instrument is needed. With its larger number of
antennas (12 vs 6), and  wide instantaneous bandwidth (16GHz), 
the NOEMA  project (NOrthern Extended Millimeter Array) 
for the IRAM Plateau de Bure Interferometer, 
will be able to produce
high quality images of a much larger sample of molecules and sources, enabling
the possibility to study the FUV chemistry in a wide
diversity of physical conditions. These data will complement the information
on the dust spectral energy distribution and gas thermal balance, which
will be provided at unprecedented sensitivity and spatial distribution
by ALMA. This will allow to probe for the first time the
 steep gradients in temperature and densities at PDR edges, as well as
the gradients in elemental abundances induced by freeze-out on grains.

Furthermore, hydrides play a key role in most chemical networks, as they
are at the root of the formation pathways of most interstellar molecules (see
the discussion by Lis et al. in this proceeding), but are however not
accessible from the ground due to the large opacity of the Earth
atmosphere at THz frequencies. With the full opening of the far infrared
and submillimeter wavelength domain by the Herschel satellite, the spatial
distribution of important hydrides such as water vapor H$_2$O, OH, CH, CH$^+$, 
NH, HF and many others including the isotopologues will be investigated. 
Although lacking spatial resolution,
and sensitivity, the pioneering detections of water by
the ISO, SWAS and ODIN satellite have demonstrated the added value of detecting
hydrides for improving the PDR models \citep{cer06,ber03,per09,gon08}. 
The launch of the Herschel Space Observatory 
in May 2009 has concretized many decades of efforts for
proposing a spatial submillimeter and far infrared telescope with 
direct detection detectors and heterodyne receivers. 
T. Phillips has been among the strongest advocates for such a
space mission. The Herschel community warmly thanks him for
this major achievement.

\acknowledgements 
M.G. and J.P.  are  indebted to T. Phillips for welcoming them in his group at
CalTech. We enjoyed working in the CSO group, the stimulating atmosphere and
Tom's insightful and enthusiastic comments on new ideas.  The CSO is
a wonderful telescope, where we learned many subtleties of 
performing heterodyne observations, from the behavior of local oscillators
and mixers, to the pointing of the antenna ...









\end{document}